\documentclass[journal=nalefd,manuscript=letter]{achemso}

\usepackage{graphicx}  
\usepackage{float}  	
\usepackage{url}     		
\usepackage{multirow}
\usepackage{dcolumn}
\usepackage{bm}
\usepackage{color}
\usepackage{amsmath}
\usepackage[version=3]{mhchem}
\usepackage{verbatim}
\usepackage[colorlinks,hyperindex]{hyperref}
	\hypersetup
		{colorlinks,	
	 	citecolor=blue,	
	 	linkcolor=blue,	
	 	urlcolor=blue,	
		}
\graphicspath{{./}} 	

\author{Wei Chen}
   \affiliation{Department of Physics and John A. Paulson School of Engineering and Applied Science, Harvard University, Cambridge, MA 02138, USA}
   \alsoaffiliation{ICQD, Hefei National Laboratory for Physical Sciences at Microscale, and Synergetic Innovation Center of Quantum Information and Quantum Physics, University of Science and Technology of China, Hefei, Anhui 230026, China} 
\author{Yuan Yang}
   \affiliation{Department of Chemistry and Chemical Biology, Harvard University, Cambridge, MA 02138, USA}
\author{Zhenyu Zhang}
   \affiliation{ICQD, Hefei National Laboratory for Physical Sciences at Microscale, and Synergetic Innovation Center of Quantum Information and Quantum Physics, University of Science and Technology of China, Hefei, Anhui 230026, China} 
\author{Efthimios Kaxiras}
   \email{kaxiras@physics.harvard.edu}
   \affiliation{Department of Physics and John A. Paulson School of Engineering and Applied Science, Harvard University, Cambridge, MA 02138, USA}

\title{Properties of In-Plane Graphene/MoS$_2$ Heterojunctions} 

\begin{document}

\begin{tocentry}

\begin{center}
\includegraphics[width=2.35 in]{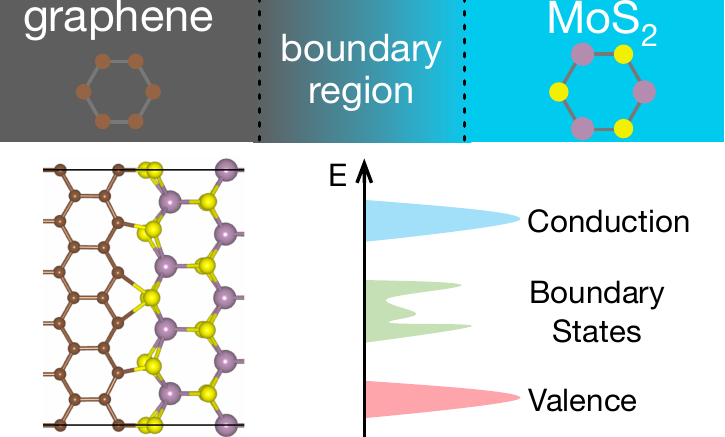}
\end{center}

\end{tocentry}

%%%%%%%%%%%%%%%%%%% >>>>>>>>>> ABSTRACT <<<<<<<<<< %%%%%%%%%%%%%%%%%%%
\begin{abstract}
The graphene/MoS$_2$ heterojunction formed by joining the two components laterally in a single plane promises to exhibit a low-resistance contact according to the Schottky-Mott rule. Here we provide an atomic-scale description of the structural, electronic, and magnetic properties of this type of junction. We first identify the energetically favorable structures in which the preference of forming C-S or C-Mo bonds at the boundary depends on the chemical conditions. We find that significant charge transfer between graphene and MoS$_2$ is localized at the boundary. We show that the abundant 1D boundary states substantially pin the Fermi level in the lateral contact between graphene and MoS$_2$, in close analogy to the effect of 2D interfacial states in the contacts between 3D materials. Furthermore, we propose specific ways in which these effects can be exploited to achieve spin-polarized currents.
\end{abstract}

\section{Introduction} \label{introduction} %%%%%%%%%%%%%%%%%%%

Two-dimensional (2D) layered materials, such as graphene, transition metal dichalcogenides (TMDCs), and black phosphorus, are exceptionally promising for realizing a broad variety of electronic devices~\cite{wang2012electronics,geim2013van,fiori2014electronics,novoselov20162d}, but there exists a fundamental challenge to their effective use, namely the large resistance of electrical contacts~\cite{allain2015electrical,xu2016contacts}. A Schottky barrier (SB) inhibits transport of carriers across the metal-semiconductor junction (MSJ), and the tunability of its height (SBH) is often quite limited due to Fermi level (FL) pinning~\cite{kurtin1969,tersoff1984schottky,tung2000chemical}. Specifically, 2D materials tend to form weak van der Waals (vdW) interactions with other 2D and 3D structures~\cite{allain2015electrical,liu2016van,ajayan2016two,zhou2016enhancing}; this introduces a gap between vdW-bonded planes of atoms across which there are no covalent or metallic bonds, which significantly degrades the contact quality by causing an \textit{additional} tunnel barrier (TB) for carriers~\cite{allain2015electrical,ajayan2016two,pierucci2016band}. Creating pure edge contacts in 2D materials, an approach that could reduce this TB~\cite{wang2013one}, is difficult between the different planar structures of these systems~\cite{allain2015electrical}.

Recently, substantial progress has been made in fabricating heterojunctions between 2D materials on a single plane, where seamless edge-to-edge covalent bonds are formed at the interface; we refer to these as ``lateral heterojunctions''. Examples include in-plane stitching of (i) one-atom-thick sheets, like graphene/hexagonal boron nitride (h-BN)~\cite{liu2013plane,liu2014heteroepitaxial}, (ii) different TMDC systems or phases~\cite{huang2014lateral,duan2014lateral,lin2014atomic,kappera2014phase,nourbakhsh2016mos}, like MoSe$_2$/WSe$_2$ or 1T/2H phase boundaries of MoS$_2$, and (iii) one-atom-thick sheet and TMDC, like graphene/MoS$_2$~\cite{ling2016parallel,zhao2016large,guimaraes2016atomically,chen2016lithography}. These systems provide a promising platform for realizing low-resistance contacts between 2D metals and 2D semiconductors~\cite{allain2015electrical,ajayan2016two,nourbakhsh2016mos}, that can be assembled into field-effect transistors (FETs) with remarkable performance~\cite{kappera2014phase,nourbakhsh2016mos,ling2016parallel,zhao2016large,guimaraes2016atomically}. To improve the device performance of 2D MSJ, a better fundamental understanding of its electronic properties and operating mechanism is needed. Recently, Yu \textit{et al}. reported a model study based on semiclassical macroscopic theory~\cite{yu2016carrier}, that suggests a highly nonlocalized charge transfer and strong reduction of FL pinning in 2D MSJs. An atomic-scale understanding of these issues is still lacking, because the properties of the 1D interface between 2D materials are \textit{governed} by the local chemical interactions at the boundary, a challenging situation due to the complexity of the boundary structure.

\begin{figure}[h]%[H]%[b!] %b!
	\begin{center}
		\includegraphics[width=5.3 in]{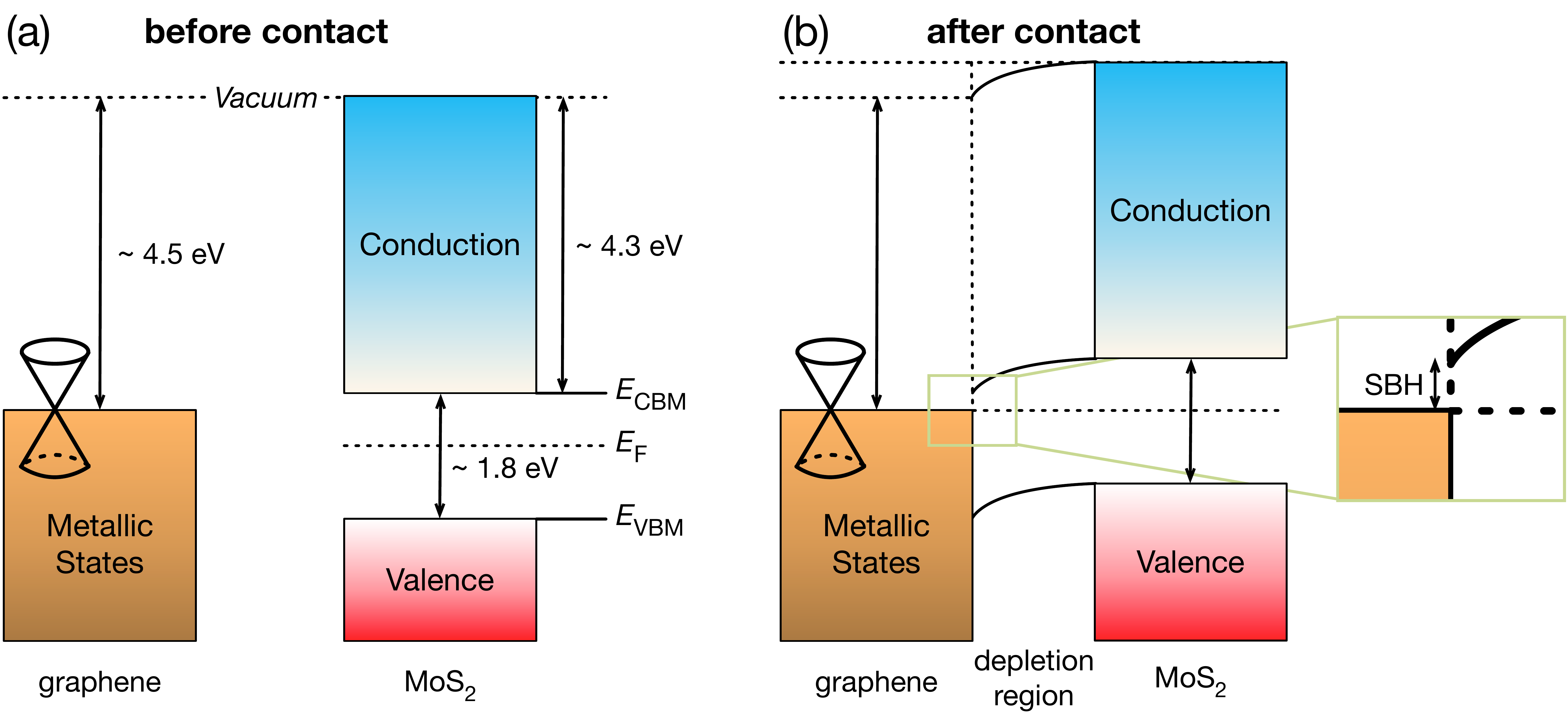}
	\end{center}
	\caption{Band alignment between undoped graphene and monolayer MoS$_2$ (a) before and (b) after contact, without considering FL pinning. The work function of monolayer graphene is about 4.5 eV~\cite{yu2009tuning,xu2012direct}. The calculated electron affinity and band gap of monolayer MoS$_2$ are consistent with previous studies~\cite{popov2012designing,chen2013tuning,gong2013band,kang2013band,liu2016van,nourbakhsh2016mos}. $E_\text{F}$ stands for Fermi energy, CBM for conduction band minimum, and VBM for valence band maximum.}
	\label{figure_band_align}
\end{figure}

Here we report a first-principles study based on density functional theory (DFT) of the interfacial properties of graphene/MoS$_2$ lateral junction. This system is of particular interest because: (a) no phase engineering is needed to create metastable structures, making the overall system stable under working conditions and its intrinsic properties unperturbed by adsorbates that induce phase transitions; (b) graphene is a low-work-function metal electrode and monolayer MoS$_2$ is generally an \textit{n}-type semiconductor~\cite{radisavljevic2011single,li2012optical,lee2012synthesis}, a combination which naturally results in a small SBH according to the Schottky-Mott rule (Fig.~\ref{figure_band_align}); (c) graphene and MoS$_2$ have significant crystallographic difference which makes it difficult to infer their interfacial structure. In the following, we identify the possible atomic structures of graphene/MoS$_2$ stitching which depend on the growth conditions. Focusing on the most stable configurations, we then analyze the charge transfer between the two sides of the boundary. Next, by calculating the energy positions and densities of the boundary states, we evaluate the strength of FL pinning effect in graphene/MoS$_2$ 2D lateral MSJs. Finally we point out how the investigated magnetic properties can be exploited for spintronic applications.

\section{Methods} \label{method} %%%%%%%%%%%%%%%%%%%

We performed the DFT calculations using the Vienna \textit{ab initio} simulation package (VASP)~\cite{kresse1996eff} with the projector-augmented wave (PAW) potentials~\cite{blochl1994pro,kresse1999from}. We employed the generalized gradient approximation parameterized by Perdew-Burke-Ernzerhof (GGA-PBE)~\cite{perdew1996gen} for the exchange-correlation functional and the DFT-D3 correction method~\cite{grimme2010consistent} for the vdW interactions, with an energy cutoff for the plane-wave basis of 450 eV. We find the lattice constants of graphene and MoS$_2$ (2H phase) to be 2.47 \AA\ and 3.16 \AA, respectively, values that are within 0.4\% (for graphene) and 0.1\% (for MoS$_2$) of experiment. We model the lateral junctions of graphene and MoS$_2$ by stitching edges of different configurations. In the supercell, the \textit{y}-axis is along the 1D boundary, with vacuum layers of sufficient thickness added in both \textit{x} and \textit{z} directions to ensure decoupling between neighboring images. The graphene edges away from the boundaries are passivated by hydrogen atoms. During structural relaxation, all the atoms are allowed to relax until the forces on them become smaller in magnitude than 0.01 eV/\AA. We use $\Gamma$-centered \textit{k}-point meshes of 1$\times$3$\times$1 for structural relaxations, 1$\times$6$\times$1 for charge density calculations, and 1$\times$12$\times$1 for density of states (DOS) calculations.

\section{Results} %%%%%%%%%%%%%%%%%%%

\subsection*{Structural dependence on growth conditions} %%%%%%%%%%%%%%%%%%%

We consider parallel stitching of the \textit{zigzag} edges of graphene and MoS$_2$, which are the predominant edge configurations found in the epitaxial growth of each material~\cite{yu2011control,lauritsen2007size,yazyev2014polycrystalline,cao2015role}. Because of the large lattice mismatch of the unit cells (2.47 \AA\ \textit{vs}. 3.16 \AA), we choose widths of each cell that are multiples of the lattice constants so that the two sides have approximately the same size; the multiples we chose are $N_{\text{gra}}=5$ and $N_{\text{MoS}_2}=4$. Graphene has a much larger Young's modulus than MoS$_2$~\cite{jiang2015graphene}, therefore the lattice length of MoS$_2$ is compressed by 2.3\% along the boundary direction, to match exactly with that of graphene.

\begin{figure}[h]%[H]%[b!] %b!
	\begin{center}
		\includegraphics[width=5.0 in]{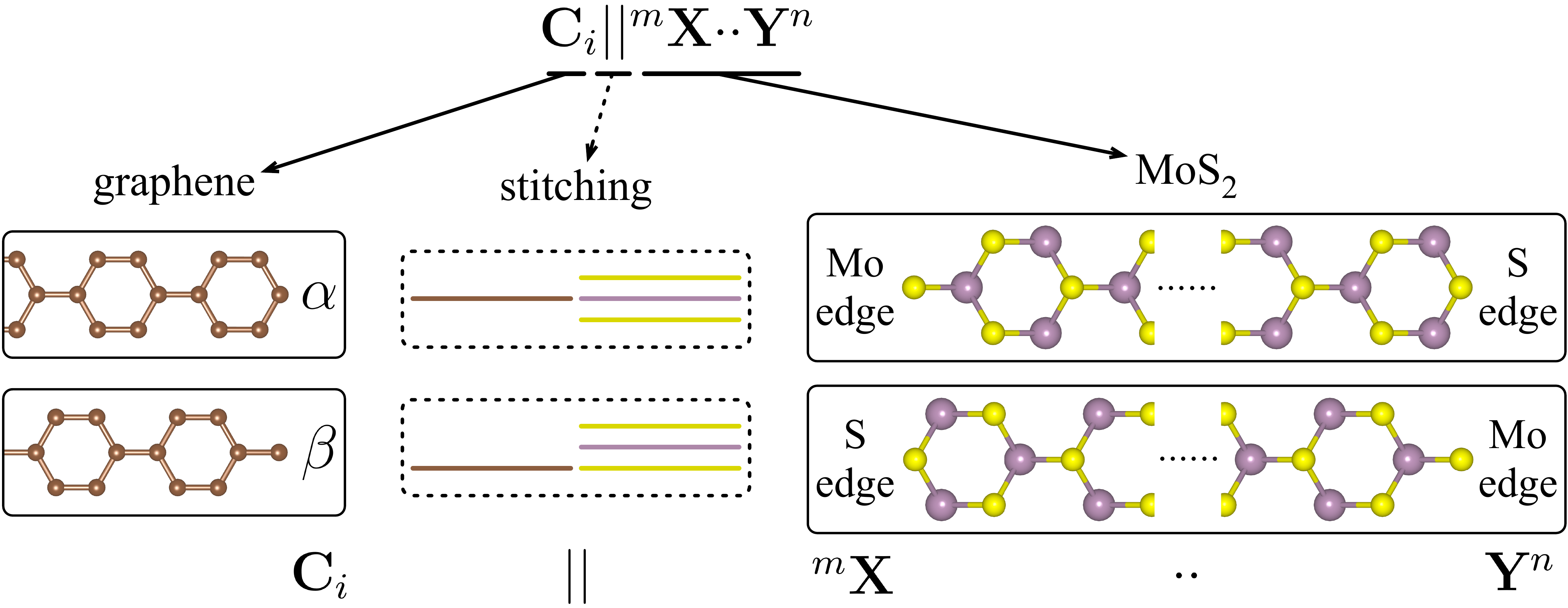}
	\end{center}
	\caption{The notation and structural phase space of explored graphene/MoS$_2$ lateral heterojunctions, consisting of two types of graphene zigzag edges $(\alpha$ and $\beta)$ and two-types of MoS$_2$ zigzag edges (Mo-edge and S-edge, each with different S passivation). The crystalline symmetry constrains the two sides of MoS$_2$ to be different types. We only show the 100\% S-passivated MoS$_2$ edges, although all the other possible concentrations~\cite{lauritsen2007size} are also investigated. To construct the boundary, graphene is initially positioned at the level of Mo plane or S plane, followed by structural relaxation.}
	\label{figure_struc_exploration}
\end{figure}

We calculated 37 different structures to find out the most stable ones. Here we use the notation of {\bf{C$_i ||^{m}$X$\cdot\cdot$Y$^{n}$}} (Fig.~\ref{figure_struc_exploration}) to represent each individual structure investigated, in which {\bf{C}} stands for carbon, $i$ indicates the type of graphene zigzag edge ($i\in\{\alpha,\beta\}$), {\bf{X}} ({\bf{Y}}) indicates the type of MoS$_2$ edge attached (unattached) to graphene as defined by its atomic termination ($\{${\bf{X}}, {\bf{Y}}$\}$=$\{${\bf{Mo}}, {\bf{S}}$\}$), $m$ and $n$ are the sulfur-passivation concentrations of MoS$_2$ edges ($m$, $n\in\{0,50,75,100\}$). For example, {\bf{C$_\alpha ||^{100}$S$\cdot\cdot$Mo$^{50}$}} represents $\alpha$-type graphene edge stitching with 100\% S-passivated S-edge of MoS$_2$, while the other side of MoS$_2$ stripe is 50\% S-passivated Mo-edge.

For each heterojunction structure the total energy $E_{\text{total}}$ is composed of energy contributions from the boundary region $E_{\text{\bf{C}}_i||^m\text{\bf{X}}}$, the MoS$_2$ edge unattached to graphene $E_{\text{\bf{Y}}^n}$, and the remaining part which is essentially equivalent in different junction configurations. We assume $E_{\text{\bf{C}}_i||^m\text{\bf{X}}}$ and $E_{\text{\bf{Y}}^n}$ are \textit{decoupled}, namely the structural variation of 
$\text{\bf{C}}_i||^m\text{\bf{X}}$ does not affect the energy of ${\text{\bf{Y}}^n}$, and vice versa. This is a reasonable approximation when the MoS$_2$ part is wide enough to eliminate interaction between the two edges. We therefore perform a \textit{two-step} search, to determine the most favorable junction structures: (i) {\bf{C$_i ||^{m}$Mo$\cdot\cdot$S$^{100}$}} and {\bf{C$_i ||^{m}$S$\cdot\cdot$Mo$^{100}$}} with different $i$, $m$ values and stitching patterns are calculated to find out the preferred boundary ($\text{\bf{C}}_i||^m\text{\bf{X}}$) configurations; (ii) using the optimal $\text{\bf{C}}_i||^m\text{\bf{X}}$ structures obtained, ${\text{\bf{Y}}^n}$ edges with varying $n$ values are compared regarding their stability.

\begin{figure}[h]%[H]%[b!] %b!
	\begin{center}
		\includegraphics[width=5.3 in]{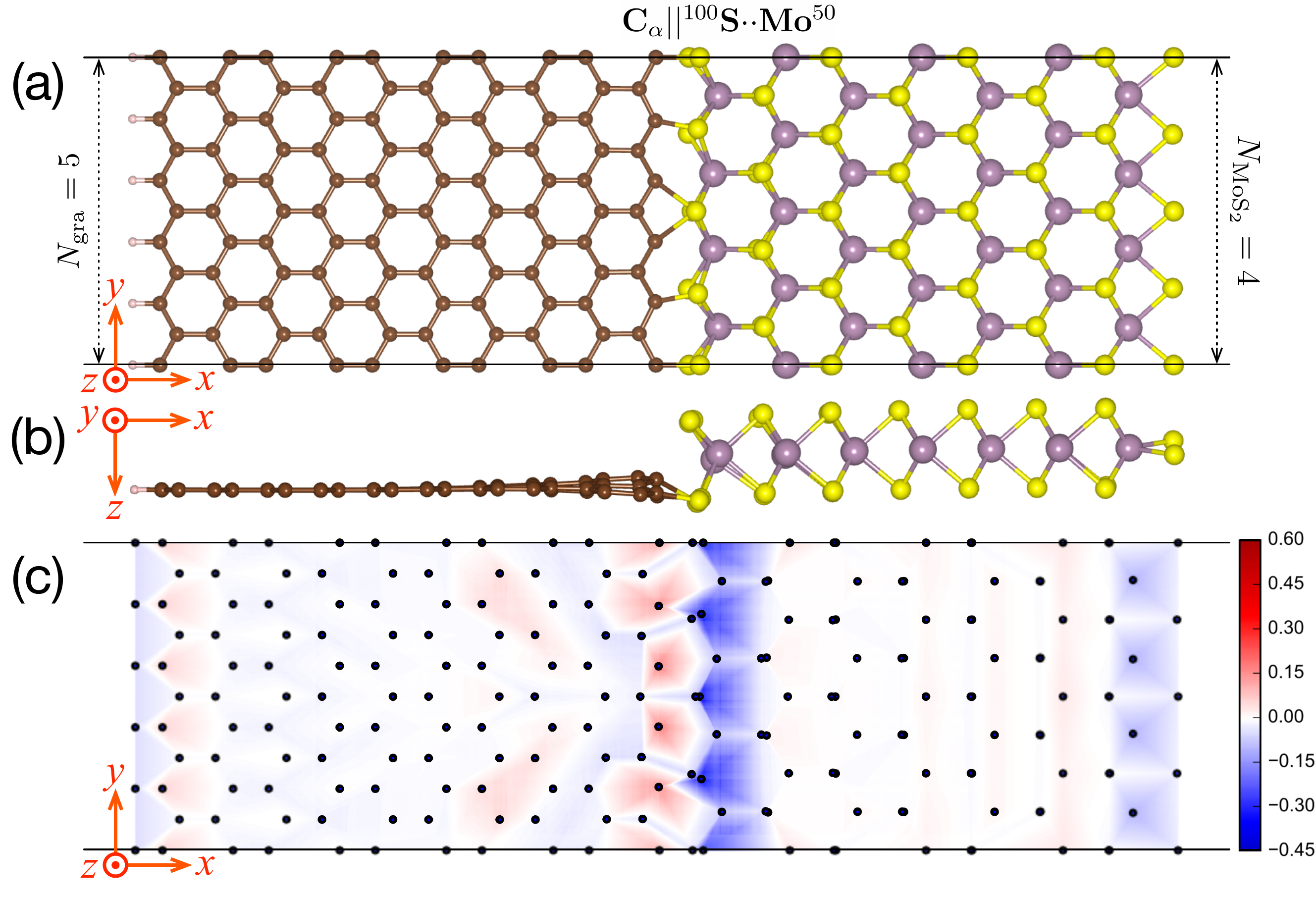}
	\end{center}
	\caption{(a) Top and (b) side views of the atomic structure and (c) top view of the Bader charge redistribution of the Phase III graphene/MoS$_2$ junction. The colorbar is in units of $e$: a positive value means gain of electronic charge, and negative means loss of electronic charge.}
	\label{structure_phase3}
\end{figure}

To compare structures with variable stoichiometry, we define the formation energy ($E_f$) per boundary length ($L$) as~\cite{bollinger2003atomic} 
\begin{equation}
\label{formation_energy}
\gamma=\frac{1}{L}E_f=\frac{1}{L}(E_{\text{total}}-n_{\text{C}}\mu_{\text{C}}-n_{\text{H}}\mu_{\text{H}}-n_{\text{Mo}}\mu_{\text{MoS}_2}-\Delta n_{\text{S}}\mu_{\text{S}}).
\end{equation}
$L=N_{\text{gra}}\cdot a_{\text{gra}}=12.34$ \AA, where $a_{\text{gra}}$ is the calculated lattice constant of graphene. $n_{\text{X}}$ and $\mu_{\text{X}}$ refer to the number of atoms in the junction and the chemical potential, respectively, of element X, except for $\mu_{\text{MoS}_2}$, which is the energy of the primitive unit cell of monolayer MoS$_2$, and $\Delta n_{\text{S}}=n_{\text{S}}-2n_{\text{Mo}}$. $\mu_{\text{C}}$ is obtained from the energy of pristine monolayer graphene, and $\mu_{\text{H}}$ from the energy of a H-passivated graphene zigzag nanoribbon (including $N_{\text{C}}$ carbon atoms) after subtracting $N_{\text{C}}\mu_{\text{C}}$. $\mu_{\text{S}}$ is a variable, whose value depends on the growth conditions, including temperature $T$, pressure $p$, etc. Assuming quasi-equilibrium growth of MoS$_2$ ($\mu_{\text{MoS}_2}=\mu_{\text{Mo}}+2\mu_\text{S}$) and no bulk Mo or S precipitation ($\mu_{\text{Mo}}<{\mu_{\text{Mo(bulk)}}}$ and $\mu_{\text{S}}<{\mu_{\text{S(bulk)}}}$)~\cite{bollinger2003atomic,cao2015role,huang2016dumbbell}, we estimate the value range of $\mu_\text{S}$ as: ($\frac{1}{2}(\mu_{\text{MoS}_2}-\mu_{\text{Mo(bulk)}})$, $\mu_{\text{S(bulk)}}$), where the upper and lower bound corresponds to the S-rich and Mo-rich growth conditions, respectively. Using the energy of body-centered cubic Mo crystal for the value of $\mu_{\text{Mo(bulk)}}$, we obtain $-$1.30 eV $<\mu_\text{S}-\mu_{\text{S(bulk)}}<0$. $\mu_{\text{S(bulk)}}$ is estimated from the energy of the crown-like S$_8$ cluster (see {\color{blue}Supplementary Fig. S1} for structures related to calculations of $\mu_{\text{H}}$, $\mu_{\text{Mo(bulk)}}$, and $\mu_{\text{S(bulk)}}$). As a result of our exhaustive search (see details in {\color{blue}Supplementary Figs. S2} and {\color{blue}S3}), we find four dominant phases, each having minimal $\gamma$ at a finite range of $\mu_{\text{S}}$ (Fig.~\ref{formation_energy}). To illustrate the salient features of the four phases indicated by the shaded regions, we show in Fig.~\ref{structure_phase3} the atomic structure of Phase III, whose stability spans over the broadest range of $\mu_{\text{S}}$ (structures of the other phases, I, II, and IV, are included in {\color{blue}Supplementary Fig. S4}).

\begin{figure}[h]%[H]%[b!] %b!
	\begin{center}
		\includegraphics[width=4.1 in]{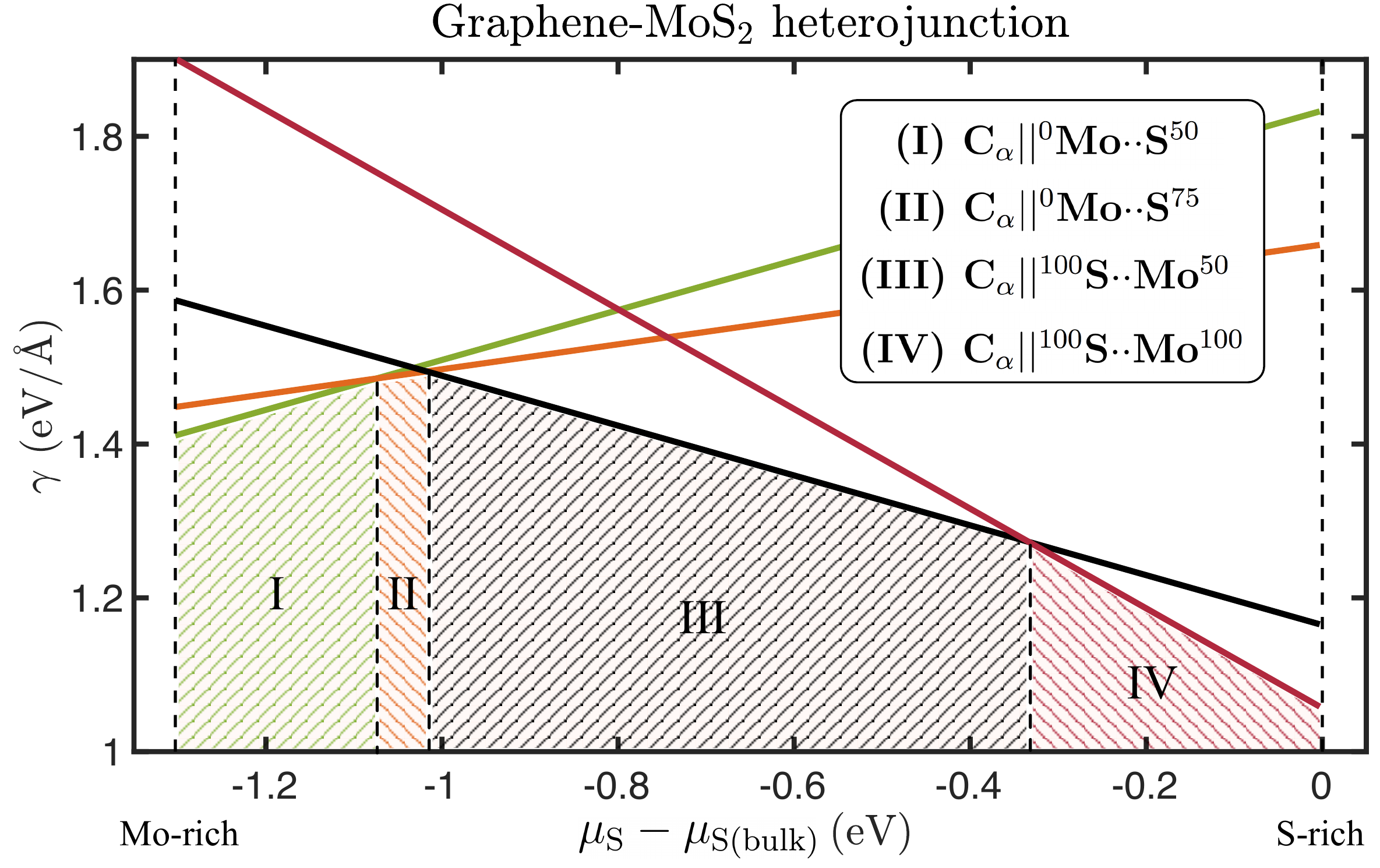}
	\end{center}
	\caption{Formation energies per unit length of the boundary of the most stable graphene/MoS$_2$ junction structures, labelled as I, II, III, and IV, as a function of S chemical potential. The atomic structure of Phase III is displayed in Fig.~\ref{structure_phase3} (see {\color{blue}Supplementary Fig. S4} for other phases).}
	\label{formation_energy}
\end{figure}

We find that, throughout the whole range of $\mu_{\text{S}}$ investigated, the \textit{$\alpha$-type graphene edge is always favored} to form interfaces with MoS$_2$. At the relatively low-$\mu_{\text{S}}$ region (close to Mo-rich conditions), the bare Mo-edge with no S passivation tends to forms Mo-C bonds at the interface, with graphene positioned at the level of Mo plane. The other side of the MoS$_2$ stripe (S-edge) is passivated by 50\% (Phase I) or 75\% S (Phase II). As $\mu_{\text{S}}$ increases, the S-edge with 100\% saturation becomes favored to interface with graphene and the graphene layer is positioned at the level of the S plane, forming S-C bonds. Due to the local strain effect and breaking of mirror-plane symmetry, there is slight deviation of carbon atoms from planarity close to the interface (Fig.~\ref{structure_phase3}(b)). As $\mu_{\text{S}}$ increases further to approach the high-$\mu_{\text{S}}$ region (close to S-rich conditions), the interfacial configuration remains intact, while the S passivation of the MoS$_2$ edge away from the interface increases from 50\% (Phase III) to 100\% (Phase IV).

\subsection*{Charge transfer effects} %%%%%%%%%%%%%%%%%%%

Having established the atomic structures of the junction, we investigate charge transfer effects at the boundary. First, we calculate the charge enclosed within the Bader volume~\cite{tang2009grid} of each atom in the primitive unit cells of graphene and MoS$_2$. The obtained Bader charges on C, Mo, and S ($Q_{\text{C}}$, $Q_{\text{Mo}}$, and $Q_{\text{S}}$) are used as reference values (in addition, $Q_{\text{H}}=1$). Second, we calculate the Bader charge on each atom (element X) in the junction structure and subtract the corresponding reference value ($Q_{\text{X}}$) to determine the amount of charge transfer ($\Delta Q$). We then extend $\Delta Q$, a relatively sparse data set, to $\Delta Q(x,y)$ on a dense mesh-grid of the $xy$-plane, with the values at the locations between atoms linearly interpolated. The resulting contour plot of the Bader charge redistribution for the Phase III junction is shown in Fig.~\ref{structure_phase3}(c) (see {\color{blue}Supplementary Fig. S5} for the other junction structures). In each of the four structural phases, substantial charge transfer is essentially \textit{localized} at the boundary and the two edges (leftmost and rightmost). Specifically, at the interface between graphene and the Mo-edge of MoS$_2$ (Phases I and II), the five C atoms forming the C-Mo bonds gain a charge of about 0.26-0.52 $e$ each, while the four Mo atoms lose a charge of about 0.12-0.15 $e$ each. When graphene forms a contact with the S-edge MoS$_2$ (Phases III and IV), the interfacial C atoms participating in the C-S bonds gain a smaller charge (about 0.14-0.22 $e$ each), while the S atoms lose more charge (about 0.35-0.38 $e$ each, see Fig.~\ref{structure_phase3}(c)). At the leftmost graphene edges, each H atom loses a charge of about 0.04-0.07 $e$, and a column of C atoms close to the edge gain a charge of about 0.1 $e$ each. At the rightmost MoS$_2$ edges, the charge redistribution is very sensitive to the edge geometry. For example, in Phase III, the rightmost column of Mo atoms lose a charge of about 0.15 $e$ per atom, and the second rightmost column of Mo gain a charge of about 0.03-0.05 $e$ per atom, while in the Phase IV, the rightmost column of S atoms lose a charge of about 0.34-0.48 $e$ per atom.
	
The Bader charge redistribution reflects how different the local bonding environment of each atom in the junction is, relative to that in the pristine monolayer graphene or MoS$_2$. The significant charge transfer observed above, indicates changes of the local bonding features at the interfaces, which strongly suggest that the electronic states of the boundary are \textit{distinct from the interior states}.

\subsection*{Boundary states and Fermi level pinning} %%%%%%%%%%%%%%%%%%%

\begin{figure}[h]%[H]%[b!] %b!
	\begin{center}
		\includegraphics[width=4.9 in]{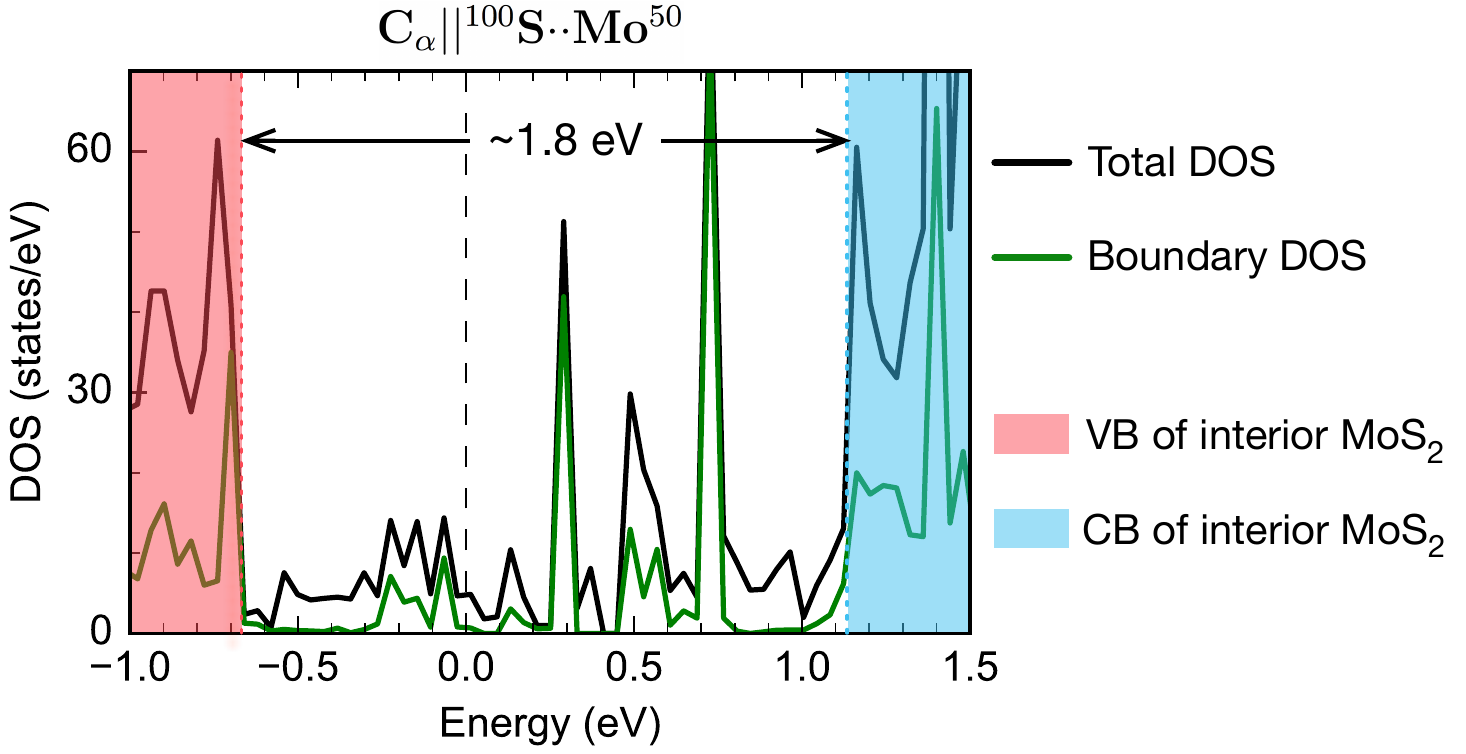}
	\end{center}
	\caption{The total and boundary DOS of the Phase III graphene/MoS$_2$ junction. The FL is set at $E_\text{F}=0$ eV; results for the other junction structures are given in {\color{blue}Supplementary Fig. S6}.}
	\label{dos_phase3}
\end{figure}

Next, we calculate the total and local density of states (DOS) of the junction, to determine the energy of the boundary states with respect to the band edges of interior bulk-like MoS$_2$. In Fig.~\ref{dos_phase3} we show the total DOS consisting of the electronic states of all the atoms in the cell, which includes 193 atoms for the Phase III structure. Based on the charge redistribution, we choose atoms that form the carbon hexagons and Mo-S hexagons closest to the interface (including 20 C, 8 Mo, and 16 S atoms), to calculate the local DOS of the boundary. A column of Mo atoms in the middle of the MoS$_2$ stripe are chosen to determine the band edges of the interior MoS$_2$ states. Limited by the computational capability (each system consists of $\sim$ 200 atoms), the calculated interior states may be more or less affected by the edge or boundary states; however, a band gap of about 1.8 eV can be clearly identified between the conduction band (CB) and valence band (VB).

There are several mid-gap states associated with the boundary of the junction. For a quantitative description, we integrate the boundary DOS over energy within the band gap of the MoS$_2$ interior states, and obtain the total number of the interfacial states to be about 10.23. In the cell, the boundary length is 1.234 nm. The linear density of interfacial states at the contact can therefore be estimated $n\approx8.3$ states/nm for the Phase III structure. Using the same approach, we calculate the values of $n$ to be about 6.3, 7.1, and 7.5 states/nm for the Phases I, II, and IV structures, respectively (see {\color{blue}Supplementary Fig. S6}).

From this analysis, we can investigate the FL pinning effect caused by the interfacial states. An earlier semiclassical study predicted the dependence of the pinning strength, defined as $S=1-{\partial\Phi_{\text{B}}}/{\partial\Phi_{\text{w}}}$, where $\Phi_{\text{B}}$ is the SBH and $\Phi_{\text{w}}$ is the metal work function, on the linear charge density of interfacial states ($n$) in a 2D lateral junction. The FL pinning in 2D systems was then estimated to be less significant~\cite{yu2016carrier}. Using the more realistic data of $n$ estimated from our first-principles calculations, $S$ is expected to be well above 0.7 for each of the four phases, close to the regime of full FL pinning. Although the calculated structures are thermodynamically the most stable ones, defects, such as point defects~\cite{PhysRevB.95.014106} and orientational disorder~\cite{chen2012suppression,liu2014unusual}, are normally inevitable at the interfacial region during actual fabrication. Such defects will contribute to the mid-gap states and therefore further enhance the FL pinning strength. 

The work function of metallic graphene can be effectively tuned by surface adsorption of molecules or by applying external electric field. As the pinning effect is expected to be significant, the FL of MoS$_2$ is locked at the level of interfacial states rather than at the FL of graphene, leading to a hardly tunable SBH at the graphene/MoS$_2$ junction. Due to its 2D nature, the large surface area for external contacts or modifications may allow graphene to have a broad range of doping level~\cite{giovannetti2008doping,heller2016theory}, making the SBH partially tunable.

The abundance of interfacial states mainly results from the dissimilarity between the two 2D materials, including chemical elements, lattice constants, thickness, etc. A possible solution to suppress the sharp disruption of each material at the interface, is to introduce an annealing process~\cite{allain2015electrical} after the formation of junctions. During this process, graphene and MoS$_2$ can dissolve into each other, forming a transition region, where the structural and electronic properties change gradually across the boundary. This is likely to reduce the existence of interfacial states, and thus weaken the FL pinning effect.

\subsection*{Magnetism and half-metallicity} %%%%%%%%%%%%%%%%%%%

Because of their unique characteristics, 2D materials have provided a new platform for exploring potential applications in spintronics~\cite{karpan2007graphite,martins2007electronic,han2014graphene,han2016perspectives,li2016two,zeng2016enhanced,cui2016contrasting}. To address this issue, we incorporate the spin degree of freedom into our investigation of the boundary states of graphene/MoS$_2$ junctions. We find that attaching bare Mo-edge MoS$_2$ to graphene (Phases I and II) completely suppresses the ground-state magnetic order of the graphene zigzag edge, and the resultant interface is non-magnetic (see {\color{blue}Supplementary Fig. S7}). Stitching of the 100\% S-passivated S-edge and graphene (Phases III and IV) leads to a weakly ferromagnetic 1D boundary, where the maximum magnetic moment, located at the C atoms, is only about 0.11 $\mu_{\rm B}$. It is therefore difficult to establish robust ferromagnetism at the interface.

\begin{figure}[h]%[H]%[b!] %b!
	\begin{center}
		\includegraphics[width=4.7 in]{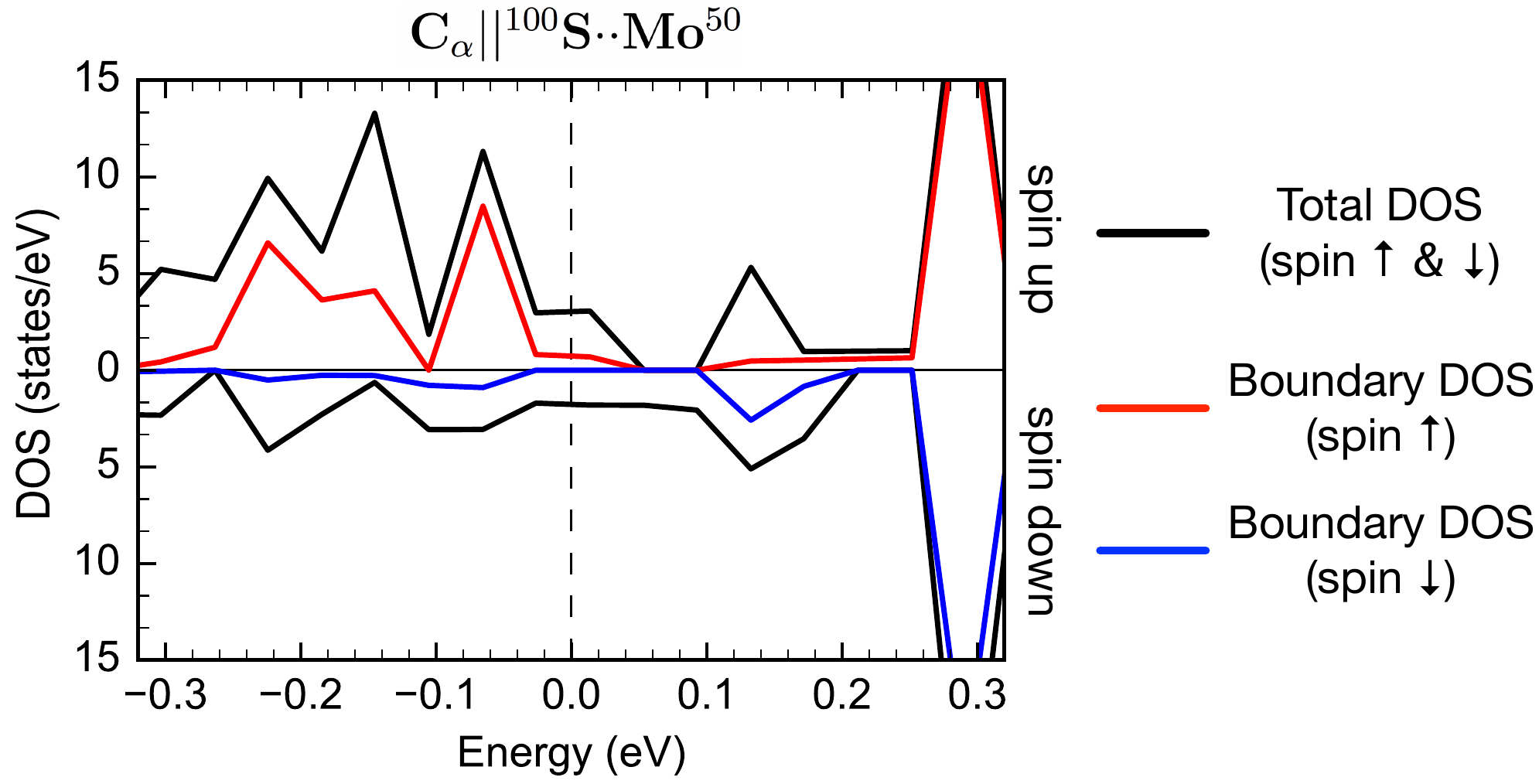}
	\end{center}
	\caption{Spin-resolved DOS of the Phase III graphene/MoS$_2$ junction. The FL is set at $E_\text{F}=0$ eV. The results for the other junction structures are shown in {\color{blue}Supplementary Fig. S9}.}
	\label{spin_dos_phase3}
\end{figure}

We plot the spin-resolved DOS for the overall structure of Phase III and its boundary region in Fig.~\ref{spin_dos_phase3}. Due to the presence of the states around FL, the whole system is metallic for both spin states (this is also seen in the band structure, {\color{blue}Supplementary Fig. S8}); however, \textit{locally}, the interfacial region is metallic for one spin while insulating for the other. The local half-metallic gap is about 0.07 eV. We analyzed the other phase structures (see {\color{blue}Supplementary Fig. S9}), and found that Phases I and II have larger local metallic gaps of about 0.16 eV, whereas Phase IV is not half-metallic. The half-metallic property therefore depends not only on the interfacial stitching structure but also on the opposite MoS$_2$ edge configuration. We hypothesize that the local half-metallicity originates from the different chemical potentials of the interface from those of the opposite graphene and MoS$_2$ edges~\cite{kan2008half}. The complexity lies in the fact that it involves the interfacial states interacting with both the graphene and the MoS$_2$ edges. The different MoS$_2$ edges have diverse localized edge states (see partial charge density distribution around the FL in {\color{blue}Supplementary Fig. S10}), leading to varying coupling strength between the interfacial states and the edge states.

\begin{figure}[h]%[H]%[b!] %b!
	\begin{center}
		\includegraphics[width=3.3in]{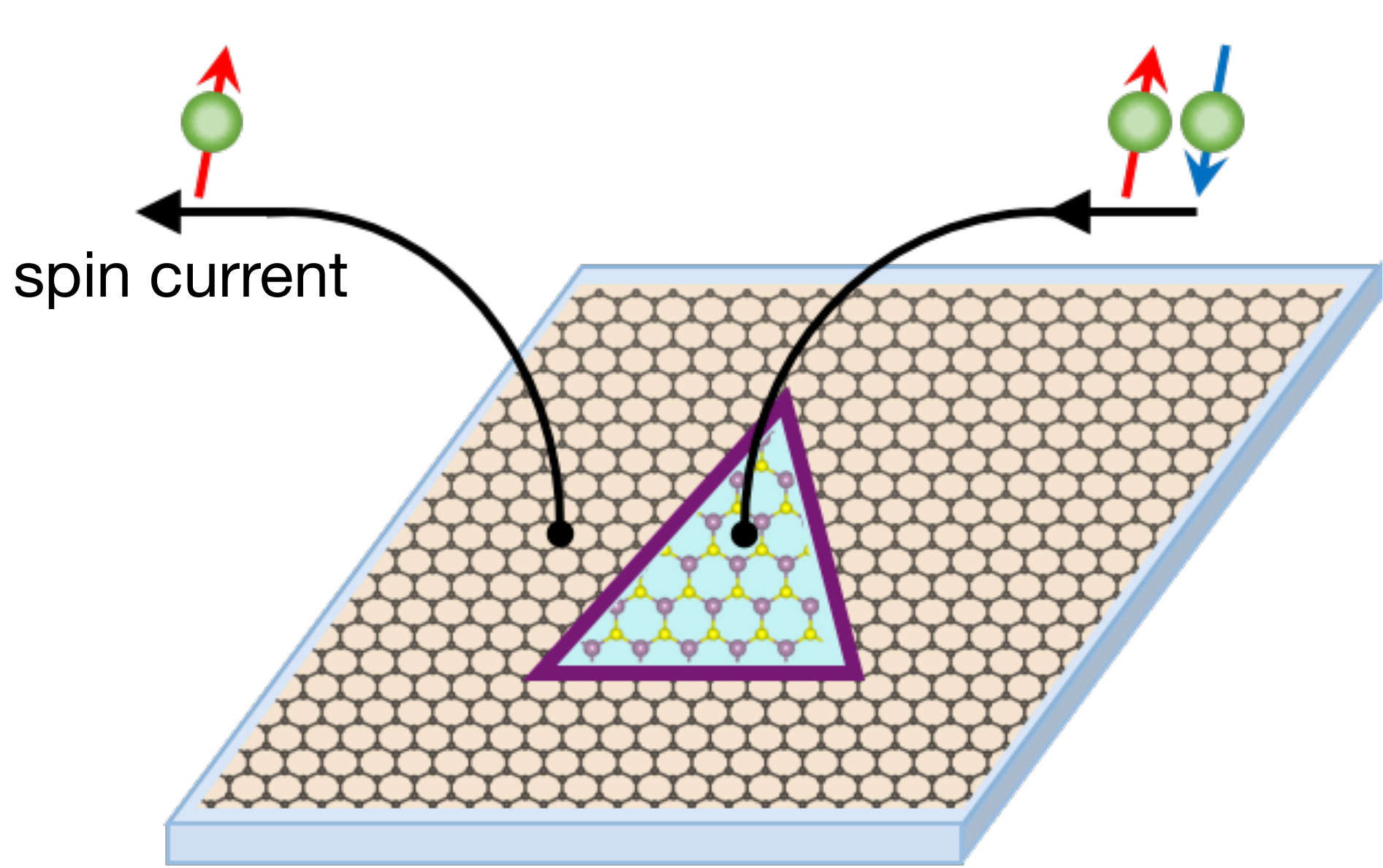}
	\end{center}
	\caption{A schematic representation of spin-polarized current generation using graphene/MoS$_2$ in-plane heterojunctions.}
	\label{spin_device}
\end{figure}

We propose that the graphene/MoS$_2$ junctions, with certain interface and edge geometries, can potentially be utilized as spin-polarized current generators~\cite{allain2015electrical}. A possible realization is shown in Fig.~\ref{spin_device}, consisting of a triangular MoS$_2$ region, embedded in the graphene layer. This device which has 3 equivalent graphene/MoS$_2$ interfaces due to the symmetry of the two components, could be realized by using graphene prepared by chemical vapor deposition (CVD)~\cite{chen2015atomistic} or mechanical exfoliation~\cite{novoselov2005two}, followed by etching of the graphene layer and epitaxial growth of MoS$_2$. When a current with both spin states is injected to MoS$_2$, only one spin polarization can tunnel through the boundary into the other component, thus exporting a spin current. 

Our results for Phases I and II indicate that, with certain structures, the graphene/MoS$_2$ junction behaves like a \textit{compensated} half metal~\cite{pickett2007half}, exhibiting zero spin moment locally at the interfacial region. The boundary can then be used to generate spin currents, without being affected by external magnetic fields. In addition, it may provide a 2D system for realizing the so-called single-spin superconductor~\cite{pickett2007half}, a novel state of both fundamental and practical significance. We emphasize that the device proposed here is an idealized prototype; the dependence of its performance on island size and edge shape, and the influence of spin-orbit coupling effects, require further systematic studies.

\section{Conclusions} %%%%%%%%%%%%%%%%%%%

In summary, we investigated the structural, electronic, and magnetic properties of the graphene/MoS$_2$ in-plane heterojunction, using first-principles calculations. We explored an extensive structural phase space to identify the atomic configurations of the junction, and found that the preference between forming C-S or C-Mo bonds at the boundary depends on growth conditions. We showed that significant charge transfer between graphene and MoS$_2$ is localized at the boundary region. The boundary contributes substantially to the mid-gap states, leading to a strong FL pinning in the junction. Finally, we demonstrated local half-metallic nature of the boundary with certain structures being compensated half metals. These findings, based on the atomic-scale understanding of the interface structure and properties, could have broad implications in the design and fabrication of purely 2D MSJs for realizing electronics with low-resistance contacts and novel spintronics devices. 

%%%%%%%%%%%%%% >>>>>>>>>> Acknowledgements <<<<<<<<<< %%%%%%%%%%%%%%
\section{Acknowledgements}
\label{sec:acknow}

We are grateful to Dr. Yu Zhu, and Shiang Fang for helpful discussions. This work was supported by the ARO MURI Award No. W911NF-14-0247, the National Natural Science Foundation of China (Grants Nos. 11504357, 61434002, 11634011), and the National Key Basic Research Program of China (Grant No. 2014CB921103). The calculations were performed at the the Oak Ridge Leadership Computing Facility (OLCF) and the National Energy Research Scientific Computing Center (NERSC) of the U.S. Department of Energy.

\bibliography{./ref}
\end{document}